\documentclass[letterpaper,english,prb, twocolumn]{revtex4-1}
\usepackage[T1]{fontenc}
\usepackage[latin9]{inputenc}
\usepackage{color}
\usepackage{babel}
\usepackage{float}
\usepackage{units}
\usepackage{amsmath}
\usepackage{amssymb}
\usepackage{graphicx}
\usepackage[unicode=true,pdfusetitle,
 bookmarks=true,bookmarksnumbered=false,bookmarksopen=false,
 breaklinks=false,pdfborder={0 0 1},backref=false,colorlinks=false]
 {hyperref}

\makeatletter

\pdfpageheight\paperheight
\pdfpagewidth\paperwidth

\usepackage{natbib}

\makeatother

\begin{document}
\global\long\def\av#1{\left\langle #1\right\rangle }

\global\long\def\abs#1{\left|#1\right|}

\title{Global Delocalization Transition in the de Moura-Lyra Model}

\author{J. P. Santos Pires}
\email{Corresponding Author: up201201453@fc.up.pt}

\address{Centro de Física das Universidades do Minho e Porto~\\
 Departamento de Física e Astronomia, Faculdade de Ciências, Universidade
do Porto, 4169-007 Porto, Portugal}

\author{N. A. Khan}

\address{Centro de Física das Universidades do Minho e Porto~\\
 Departamento de Física e Astronomia, Faculdade de Ciências, Universidade
do Porto, 4169-007 Porto, Portugal}

\author{J. M. Viana Parente Lopes}

\address{Centro de Física das Universidades do Minho e Porto~\\
 Departamento de Física e Astronomia, Faculdade de Ciências, Universidade
do Porto, 4169-007 Porto, Portugal}

\author{J. M. B. Lopes dos Santos}

\address{Centro de Física das Universidades do Minho e Porto~\\
 Departamento de Física e Astronomia, Faculdade de Ciências, Universidade
do Porto, 4169-007 Porto, Portugal}

\date{\today}
\begin{abstract}
The possibility of having a delocalization transition in the 1D de
Moura-Lyra class of models (having a power-spectrum $\propto q^{-\alpha}$)
has been the object of a long standing discussion in the literature.
In this paper, we report the first numerical evidences that such a
transition happens at $\alpha=1$, where the localization length (measured
from the scaling of the conductance) is shown to diverge as $(1-\alpha)^{-1}$.
The persistent finite-size scaling of the data is shown to be caused
by a very slow convergence of the nearest-neighbor correlator to its
infinite-size limit, and controlled by the choice of a proper scaling
parameter. Our results for these models are consistent with a localization
of eigensta\textcolor{black}{tes that is driven by a persistent small-scale
noise, which vanishes as $\alpha\to1^{-}$. This} interpretation in
confirmed by analytical perturbative calculations which are built
on previous work. Finally, the nature of the delocalization transition
is discussed and the conclusions are illustrated by numerical work
done in the $\alpha>1$ regime.
\end{abstract}
\maketitle

\section{Introduction}

It is an established fact that all eigenstates of an one-dimensional
hamiltonian usually become exponentially localized in the presence
of a random potential\,\citep{Mott1961}. Physically, this feature
translates into a peculiar behavior of the system's transport properties.
Namely, the typical ($T=0K$) conductance over a disorder ensemble
scales exponentially to zero as the length is increased, i.e. $G_{\text{Typ}}\propto\exp\left(-\nicefrac{L}{\xi}\right)$.\,\citep{lee_disordered_1985,Izrailev2012}
The characteristic parameter $\xi$ is called the localization length
and can be related to the characteristic size of the system's eigenwavefunctions
or, equivalently, to the real-space decay length of the single-particle
Green's function.\,\citep{thouless_relation_1972}

For one-dimensional models with uncorrelated disorder, the existence
of localized states at all energies was settled long ago, both by
rigorous analytical methods\,\citep{ziman_localization_1968,ishii_localization_1973}
and numerical simulations\,\citep{andereck_numerical_1980,pichard_one-dimensional_1986}.
On the contrary, the role played by space-correlations in the disordered
potential remains unclear. The early attempts to deal with correlated
disorder models\,\citep{johnston_localization_1986} did not show
any qualitative differences in the physics of the system, other than
by changing the specific values of the localization length. Later
on, more sophisticated models were invented in which the existence
of delocalized eigenstates was revealed. An important example is the
so called \emph{random dimer model} (and its generalizations\,\citep{Dunlap1990,PHILLIPS1805}),\emph{
}where the random placement of different dimers across a chain generates
delocalized states at isolated energies.

Despite having extended states, models like the ones described above
do not give rise to true mobility edges in the thermodynamic limit.
The first evidence that such a feature could appear in 1D stemmed
from the study of \emph{self-affine }potentials pioneered by de Moura
and Lyra\,\citep{deMoura1998}. They defined a random potential with
a power-spectrum decaying as $q^{-\alpha}$, which is known to occur
naturally in the assembly of biological macromolecules, such as DNA\citep{Peng1992}.
For $\alpha\approx0,$ one recovers the fully localized uncorrelated
model, while in the opposite limit ($\alpha\rightarrow+\infty$) the
potential becomes ordered and there is a complete delocalization.
By numerically studying the Lyapunov exponent as a function of $\alpha$,
de Moura and Lyra concluded that an Anderson transition happens at
$\alpha=2$, followed by the emergence of a mobility edge in the spectrum.
These results were contested\,\citep{kantelhardt_comment_2000} on
the basis of the ill-defined thermodynamic limit in these potentials.
Eventually, it was understood that the presumed transition is an artifact
of the anomalous scaling of the Lyapunov exponent, due to the non-stationarity
of the potential for any $\alpha\geq1$.\,\citep{russ_localization_2001,BUNDE2000151}
This reasoning was reexamined some years later by G. M. Petersen et
al\,\citep{Petersen2013}, who reaffirmed the appearance of a mobility
edge at $\alpha=2$ and connected it to the increasing infinite-range
anticorrelations of the disordered potential.

Despite the current knowledge about the non-stationary sector of the
de Moura-Lyra model, little attention has been given to the cases
when $\alpha\in\left[0,1\right[$. The purpose of this paper is to
fill this gap by studying the behavior of the localization length
as $\alpha\rightarrow1^{-}$, where the model goes from a truly disordered
stationary potential to a non-stationary one (where the states are
conjectured to be extended). More precisely, we report the first clear
observation of a delocalization phase transition, happening at $\alpha=1$
for all energies, without generating a mobility edge. Moreover, a
detailed analysis of the disorder's real-space correlations for $\alpha<1$
is done which, besides clarifying the origin of the persistent finite-size
scaling in the measured localization length, also shows that the most
relevant feature for the localization of the eigenstates is the magnitude
of very short-scale disorder, rather than the power-law tails of the
correlator. Anticipating the results, we state that our numerical
and analytical definitely settle the issue in favor of the full delocalization
for any $\alpha\geq1$ and, at same time, clearly invalidates the
use of the de Moura-Lyra model to study the effect of algebraic correlations
in the disordered potential, even when it is stationary.

The remaining text is organized as follows: In sect.\,\ref{ModelHam_DisStat},
we recover the definition of the de Moura-Lyra disorder model, focusing
on the calculation of its statistical properties in the thermodynamic
limit, while showing the relevance of finite-size deviations close
to the transition point; In Sect.\,\ref{Perturbative-Expression-for_LocLength},
we apply the generalized Thouless formula\citep{Izrailev1999} to
calculate perturbatively the localization length as a function of
$\alpha$. In Sect.\,\ref{Localization-Length-From_LandCond}, the
Landauer conductance of many disordered samples is numerically calculated,
and the linear scaling of its typical value is used to determine the
localization length for several values of $\alpha$. In Sect.\,\ref{CriticalBehavior},
the finite size scaling of the localization length is analyzed and
a perfect collapse of the data is accomplished in the perturbative
regime. In Sect.\,\ref{Conclusions}, we recap all the results and
sum up our conclusions.

\section{\label{ModelHam_DisStat}Model and Disorder Statistics}

The focus of this work is on the localization phenomena happening
in nearest-neighbor tight-binding chains with an on-site disordered
potential having long-ranged space correlations. The respective Hamiltonian
is\vspace{-0.5cm}

\begin{equation}
\mathcal{H}=\sum_{n=0}^{L}\varepsilon_{n}c_{n}^{\dagger}c_{n}-\sum_{n=0}^{L-1}\left(c_{n}^{\dagger}c_{n+1}+c_{n+1}^{\dagger}c_{n}\right),\label{Hamiltonian}
\end{equation}
where $\varepsilon_{n}$ stands for on-site values of the disordered
potential (in units of the hopping).\vspace{-0.5cm}

\subsection{Disorder Statistics in the Thermodynamic Limit}

\vspace{-0.2cm}

In order to generate a correlated random potential, we employ the
well-known \emph{Inverse Fourier Transform Method} (IFTM)\,\citep{deMoura1998,Petersen2013,Khan2019}
meaning that the disorder profile is defined as the Fourier sum\,\footnote{The $q=0$ mode is an uniform potential contribution which may always
be neglected. All the summations over $q$ in this paper are to be
understood as summations over the allowed wavenumbers inside the First
Brillouin Zone of an one-dimensional periodic chain, i.e. $q=2\pi n/L$
with $n\in\{-\frac{L}{2},...,-1,0,1,...,\frac{L}{2}\}$.} \vspace{-0.4cm}

\begin{equation}
\varepsilon_{n}=\sum_{q\neq0}V(q)e^{iqn+i\phi_{q}},\label{IFTPotential}
\end{equation}
where $\phi_{q}$ are statistically independent random phases, obeying
the reality constraint, $\phi_{q}=-\phi_{-q}$. For the present purposes,
we are interested on the special case of the de Moura-Lyra potential,
where $V(q)=\mathcal{A}\left(\alpha\right)\abs q^{-\frac{\alpha}{2}}$,
and focus on the $0\leq\alpha<1$ sector, in which the thermodynamic
limit does not suffer from the mathematical pathologies of the $\alpha\geq1$
cases\,\citep{Petersen2013,Khan2019}. 

By construction, the ensemble average of the on-site energies is zero,
while the local variance is seen to be site-independent and equal
to \vspace{-0.4cm}

\begin{equation}
\av{\varepsilon_{n}^{2}}=2\mathcal{A}\left(\alpha\right)^{2}\sum_{q>0}q^{-\alpha}\xrightarrow[L\rightarrow\infty]{}\frac{L\mathcal{A}\left(\alpha\right)^{2}}{\pi^{\alpha}\left(1-\alpha\right)}.
\end{equation}
By fixing the variance to $\sigma_{\varepsilon}^{2}$, we also fix
the normalization constant to be $\mathcal{A}(\alpha)=\sigma_{\varepsilon}\sqrt{\left(1-\alpha\right)\pi^{\alpha}/L}$.
Similarly, we may calculate the normalized two-point correlator of
$\varepsilon_{n}$, which yields\vspace{-0.4cm}

\begin{figure}[H]
\begin{centering}
\includegraphics[width=8.85cm]{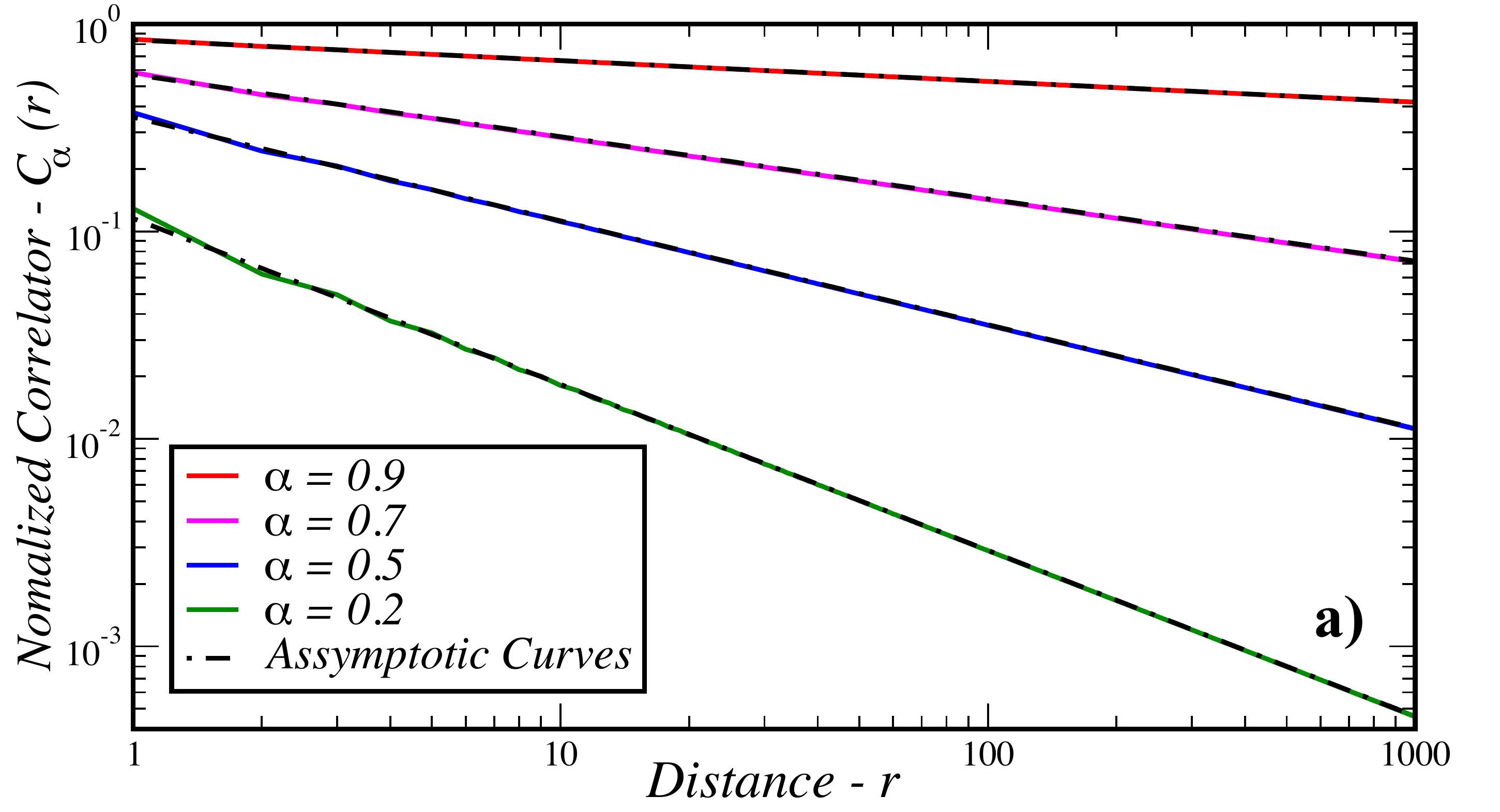}
\par\end{centering}
\begin{centering}
\includegraphics[scale=0.3]{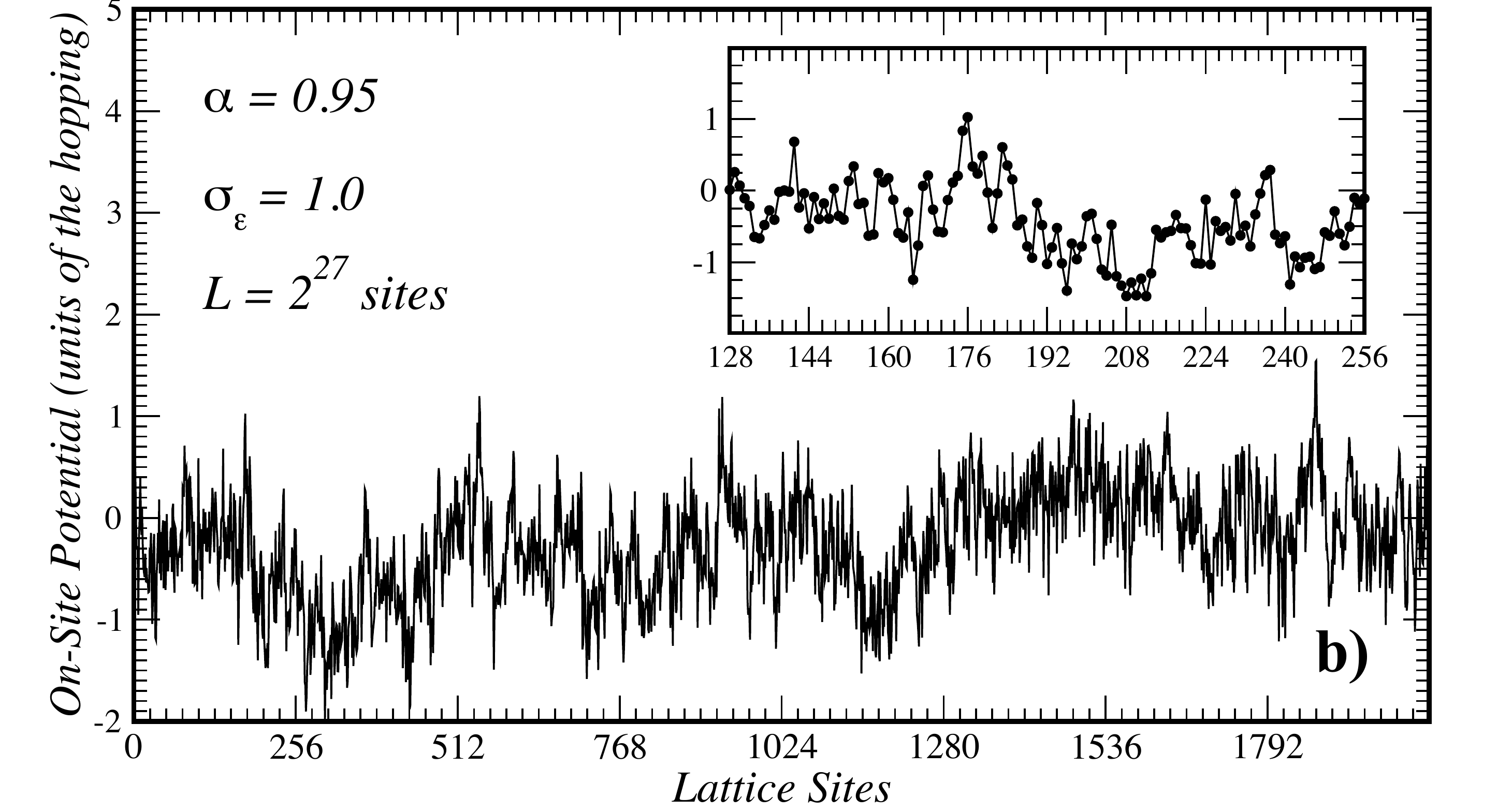}
\par\end{centering}
\centering{}\caption{\label{fig:1}\textbf{a)} Plots of the normalized correlation function
for the de Moura-Lyra model, in the limit $L\rightarrow+\infty$.
The black dashed lines are the asymptotic expressions $\mathcal{C}_{\alpha}^{\infty}(r)\propto r^{\alpha-1}$;\textbf{
b)} An example of the de Moura-Lyra disorder profile (including a
zoom in the inset). (color online)}
\end{figure}

{\small{}
\begin{align}
\mathcal{C}_{\alpha}^{\infty}(r) & =\frac{\av{\varepsilon_{n}\varepsilon_{n+r}}}{\sigma_{\varepsilon}^{2}}=\frac{1-\alpha}{\pi^{1-\alpha}}\int_{0}^{\pi}\frac{\cos qr}{q^{\alpha}}dq\label{Correlator_ThermodynamicLimit}\\
 & \qquad\qquad\qquad=\;_{1}F_{2}\left(\frac{1-\alpha}{2},\frac{1}{2},\frac{3-\alpha}{2},-\frac{\pi^{2}r^{2}}{4}\right).\nonumber 
\end{align}
}Despite being a complicated Hypergeometric function, $\mathcal{C}_{\alpha}^{\infty}(r)$
has a rather simple asymptotic expansion in $r$ (i.e. with $1\ll r\ll L$),
which shows the correlations falling-off as $r^{\alpha-1}$\,\footnote{The limit $\alpha\rightarrow0^{+}$ is somewhat special, since $\mathcal{C}_{\alpha}(r)=0$
for any $r\neq0$, recovering the usual uncorrelated Anderson disorder.}, i.e.\vspace{-0.6cm}

\begin{equation}
\mathcal{C}_{\alpha}^{\infty}\left(r\right)=\frac{\Gamma\left(\frac{1-\alpha}{2}\right)}{\Gamma\left(\frac{\alpha}{2}\right)}\frac{\pi^{\alpha-\frac{1}{2}}\left(1-\alpha\right)}{2^{\alpha}r^{1-\alpha}}+\mathcal{O}\left(\frac{1}{r^{2}}\right).\label{eq:CorrelatorAssymptotic}
\end{equation}
The plot presented in Fig.\,\ref{fig:1}\,\textbf{a)} shows that
this algebraic behavior sets in after only a few lattice spacings. 

Other than the slow decay of the correlations at large distances,
one also notices that there is a very sharp uncorrelation across a
single bond. This features disappear as $L\rightarrow\infty$ and
$\alpha\rightarrow1^{-}$ (by this order), but for a finite system
and $\alpha<1$ it is enough to generate a small-scale random noise
with an amplitude of the order of $\sigma_{\varepsilon}$ (e.g. see
Fig.\,\ref{fig:1}\,\textbf{b)}). 

\vspace{-0.9cm}

\subsection{Finite Size Effects\vspace{-0.2cm}}

In a finite sample of size $L$, an appropriate measure for the streng\textcolor{black}{th
of this short-distance noise component i}s the (squared) \emph{Normalized
Single-Bond Discontinuity} (NSBD) parameter, defined as\vspace{-0.6cm}

\begin{equation}
\mathcal{D}_{\alpha}^{L}=\frac{\av{\left(\varepsilon_{n}-\varepsilon_{n+1}\right)^{2}}}{2\sigma_{\varepsilon}^{2}}=1-\mathcal{C}_{\alpha}^{L}(1).\label{NSBD}
\end{equation}
This parameter roughly measures the dispersion of on-site energies,
relative to the previous value. In the $L\to\infty$ limit, by expanding
the function of Eq.~\ref{Correlator_ThermodynamicLimit}, this quantity
is shown to be proportional to $(1-\alpha)$ in the limit $\alpha\to1^{-}$.
On the other hand, simulated finite systems have $L$-dependent space
correlators given by\vspace{-0.45cm}

\begin{equation}
\mathcal{C}_{\alpha}^{L}(r)=\frac{\sum_{i=1}^{\nicefrac{L}{2}}i^{-\alpha}\cos\frac{2\pi ir}{L}}{\sum_{i=1}^{\nicefrac{L}{2}}i^{-\alpha}},
\end{equation}
instead of Eq.\,\ref{Correlator_ThermodynamicLimit}.

\begin{figure}[H]
\includegraphics[scale=0.4]{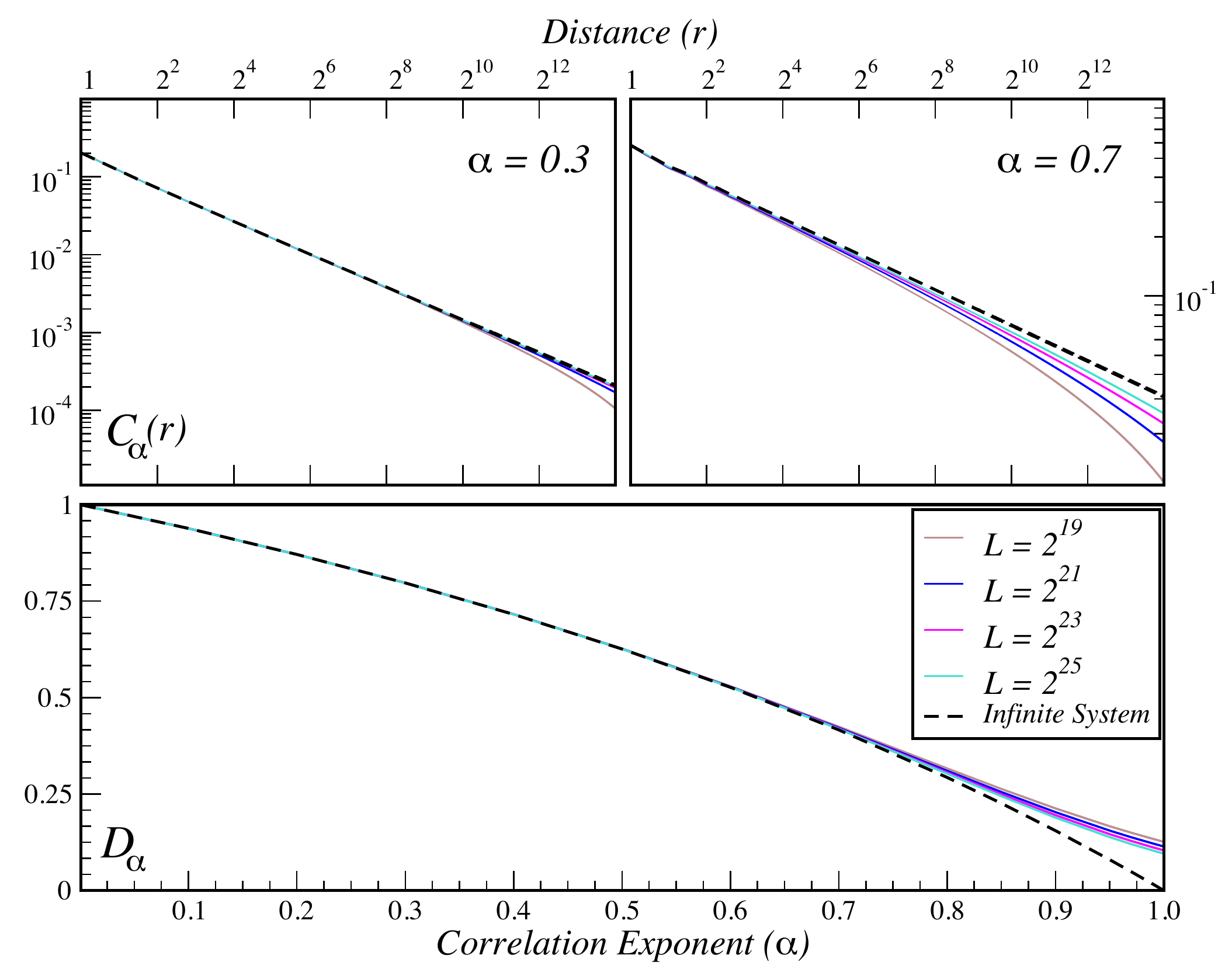}

\caption{\label{fig:2}Examples of the finite-size effects on the correlator
$\mathcal{C}_{\alpha}^{L}(r)$ (upper panels) and the NSBD parameter
$\mathcal{D}_{\alpha}^{L}$ (lower panel). The dashed lines stand
for the curves obtained in the thermodynamic limit ($L\to\infty$).
(color online)}
\end{figure}

In the Fig.\,\ref{fig:2}, we compare the exact correlator and NSBD
to the expression of Eq.\,\ref{Correlator_ThermodynamicLimit} for
different sample sizes, being quite evident the very slow convergence
towards the thermodynamic limit value, especially for values of $\alpha$
close to $1$. This very slow convergence of the NSBD to its limiting
value is shown to cause a persistent finite-size scaling of the calculated
localization length. This is also the reason for the persistence of
the small-scale noise, even when $\alpha$ is very close to $1$,
as seen in Fig.\,\ref{fig:1}\,\textbf{b).}

\section{Perturbative Expression for the Localization Length\label{Perturbative-Expression-for_LocLength}\vspace{-0.2cm}}

In the weak disordered regime, one can usually obtain analytical expressions
for the localization length, as a function of energy. In fact, F.
M. Izrailev\,\citep{Izrailev1999} derived a generalized Thouless
formula which allows the calculation of $\xi$ for arbitrary space-correlated
disordered potentials, in first order on the local variance $\sigma_{\varepsilon}^{2}$.
In the thermodynamic limit, it reads:\vspace{-0.4cm}

\begin{align}
\xi^{-1} & =\frac{\sigma_{\varepsilon}^{2}}{8\sin^{2}\left(k\right)}\left\{ 1+2\sum_{r=1}^{\infty}\mathcal{C}_{\alpha}^{\infty}\left(r\right)\cos\left(2kr\right)\right\} \label{eq:IzrailevFormula}\\
 & \qquad\qquad\qquad\qquad=\frac{\sigma_{\varepsilon}^{2}}{8\sin^{2}\left(k\right)}\sum_{r=-\infty}^{\infty}\mathcal{C}_{\alpha}^{\infty}\left(r\right)e^{i2kr},\nonumber 
\end{align}
where $k=\arccos\left(E/2\right)$ is the (positive) wavenumber associated
to an unperturbed band energy $E$. In order to apply Eq.\,\ref{eq:IzrailevFormula}
to our disorder model, we must plug in the correlator of Eq.\,\ref{Correlator_ThermodynamicLimit},
yielding\vspace{-0.5cm}

\begin{align}
\xi^{-1} & =\frac{\pi^{\alpha-1}\sigma_{\varepsilon}^{2}}{16\sin^{2}\left(k\right)}\left(1-\alpha\right)\int_{-\pi}^{\pi}\sum_{r=-\infty}^{\infty}\frac{e^{i\left(2k-q\right)r}}{\abs q^{\alpha}}dq\label{eq:IzrailevFormula-1}\\
 & \qquad\qquad\qquad\qquad\qquad\qquad=\frac{\pi^{\alpha}\sigma_{\varepsilon}^{2}}{8\sin^{2}\left(k\right)}\frac{\left(1-\alpha\right)}{\abs{2k}^{\alpha}}.\nonumber 
\end{align}

Finally, if we express it in terms of the energy, we get to the final
expression:\vspace{-0.6cm}

\begin{equation}
\xi=\frac{2}{\left(1-\alpha\right)\sigma_{\varepsilon}^{2}}\left(4-E^{2}\right)\left[\frac{2}{\pi}\arccos\left(\frac{E}{2}\right)\right]^{\alpha}.\label{LocalizationLenghtFormula}
\end{equation}

From Eq.\,\ref{LocalizationLenghtFormula} it is evident that for
any value of the band energy, the localization length is finite. However,
as $\alpha\rightarrow1^{-}$ this length diverges as $\left(1-\alpha\right)^{-1}$,
signaling the existence of a global delocalization transition at that
point, i.e without generating any mobility edge.\vspace{-0.5cm}

\section{Localization Length From the Landauer Conductance\label{Localization-Length-From_LandCond}}

The existence of a finite localization length $\xi$ leads to a quantum
conductance which has a self-averaging log-normal statistics over
the ensemble and a typical value that scales exponentially to zero
with $\nicefrac{L}{\xi}$. Hence, in order to measure it, we made
use of the linearized Landauer Formula\,\citep{landauer_electrical_1970,caroli_direct_1971,meir_landauer_1992,wimmer_quantum_2009}
for the conductance (of a two-terminal device):\vspace{-0.8cm}

\begin{equation}
G\left(E_{F}\right)=\frac{e^{2}}{h}\left(4-E_{F}^{2}\right)\abs{\mathcal{G}_{L_{S}+1,0}^{r}(E_{F})}^{2},\label{eq:Linear Conductance}
\end{equation}
where $L_{S}$ is the number of sites in the sample, $E_{F}$ is the
Fermi energy and $\mathcal{G}^{r}$ stands for the retarded real-space
Green's function. For a given sample of disorder, the latter may be
calculated using the\textbf{ }\emph{Recursive Green Function Method}\,\citep{MacKinnon1985}
with the exact surface Green's functions of the leads as boundary
conditions. 

\begin{figure}[H]
\begin{raggedleft}
\includegraphics[width=8.4cm]{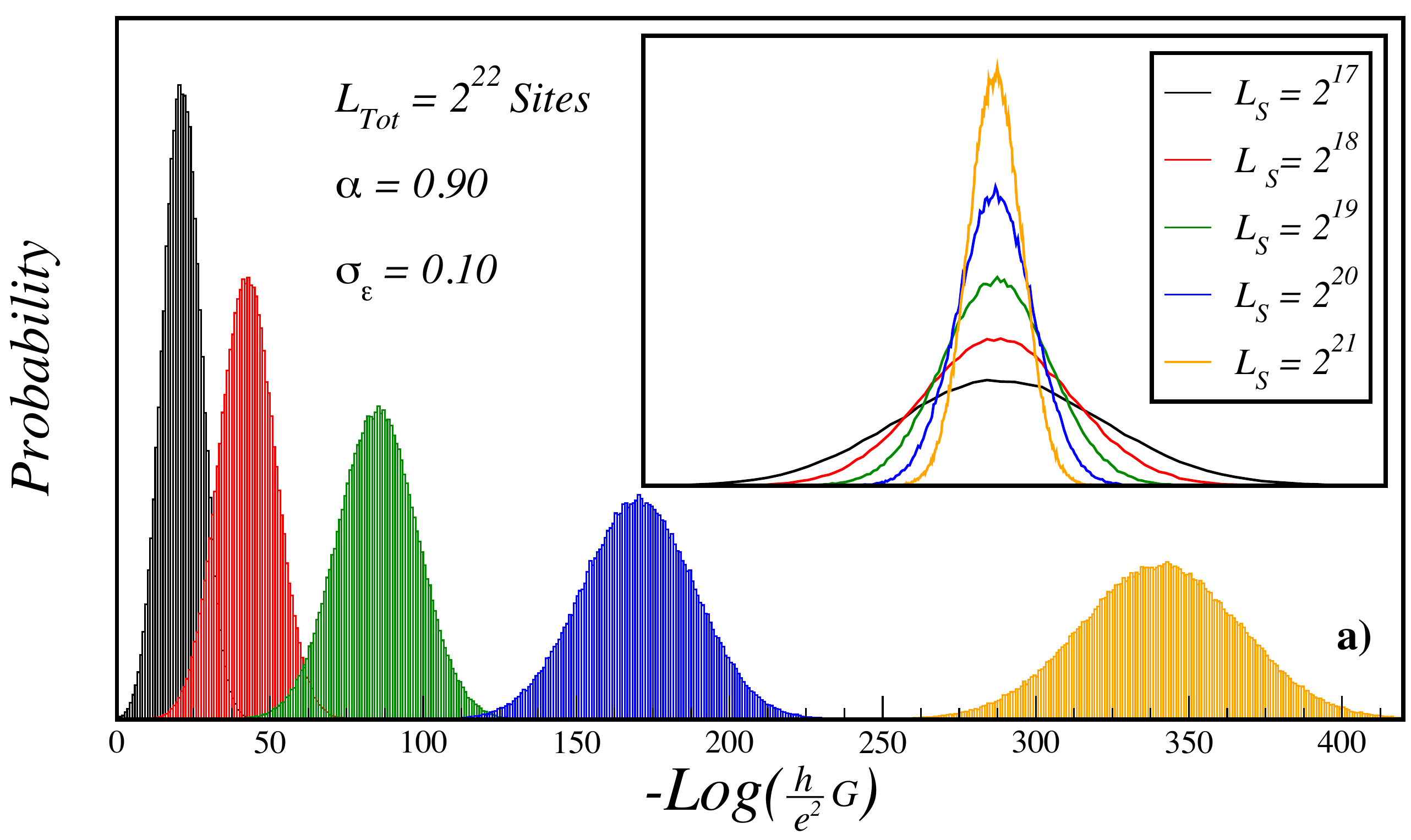}\,\,
\par\end{raggedleft}
\begin{raggedleft}
\includegraphics[width=8.5cm]{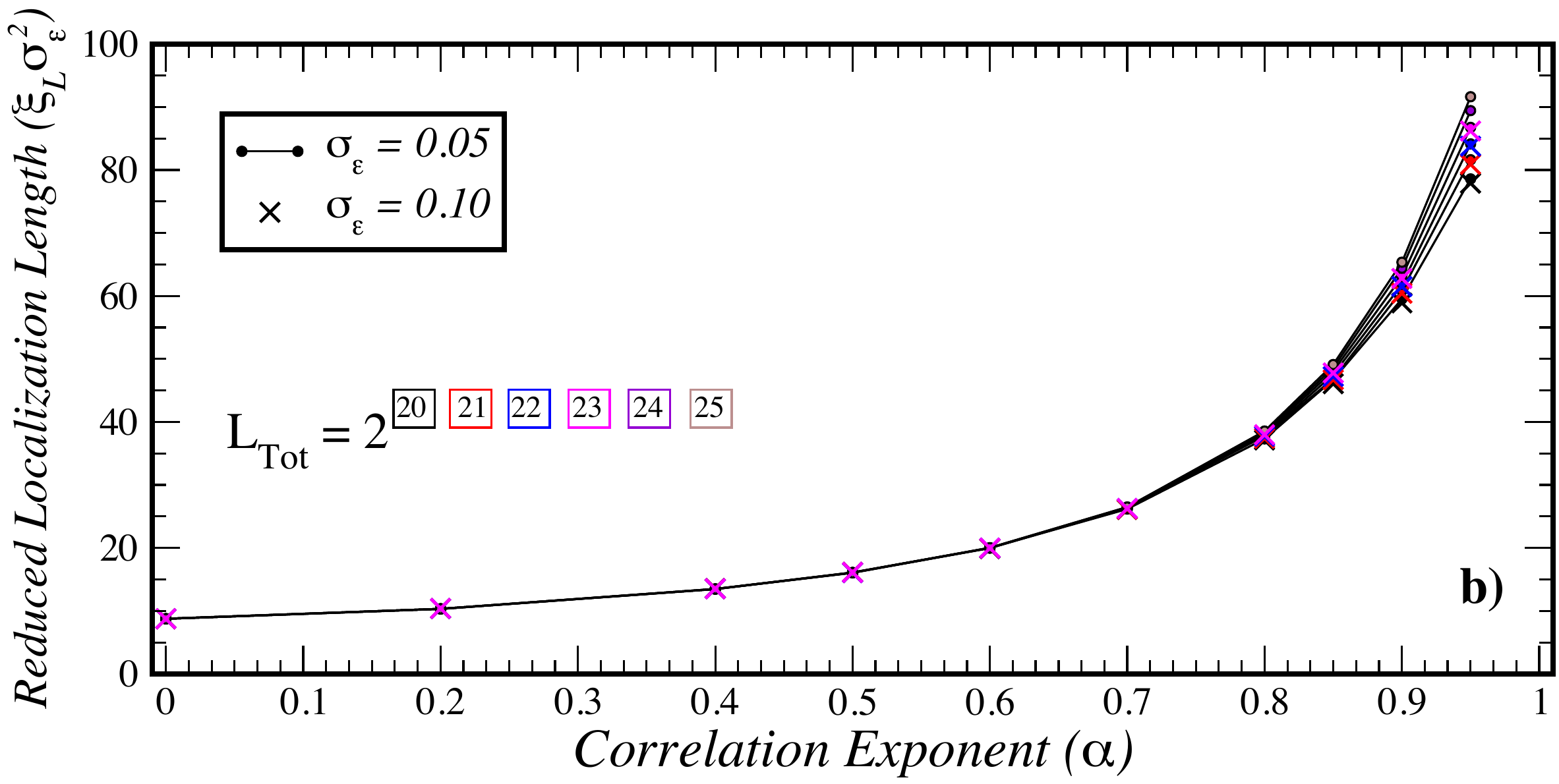}\,\,
\par\end{raggedleft}
\caption{\label{Example_histogram}\textbf{a)} Example of the histograms obtained
for the $\log\left(\frac{h}{e^{2}}G\right)$ of subchains drawn from
$10^{5}$ (independent) disordered samples with $L_{\text{Tot}}=2^{22}$
sites, $\sigma_{\varepsilon}=0.1$ and $\alpha=0.90$. In the inset,
we have the same histograms rescaled by corresponding subchain size,
making clear that this quantity is self-averaging and with an average
which scales linearly with $L_{S}$. \textbf{b) }Plot of the localization
length as a function of $\alpha$ for different values of $L_{\text{Tot}}$
(notice the color code). (color online)}
 
\end{figure}

In all our calculations, we considered the disordered samples to be
different sized pieces (subchains of size $L_{S}$) cut from independently
generated potentials with $L_{\text{Tot}}$ sites. Then, $L_{\text{Tot}}$
was increased in order to approach the thermodynamic limit of the
disorder statistics. In Fig.\,\ref{Example_histogram}\,a), we illustrate
the results with an example, from where the referred features of localization
are very evident. This same procedure was then repeated for different
values of $\alpha$, $\sigma_{\varepsilon}$ and $L_{\text{Tot}}$,
and the localization length was obtained from the inverse slope of
a linear fit to the $\av{\log\left(\frac{h}{e^{2}}G\right)}$ as function
of $L_{S}$. Some of those results are presented in Fig.\,\ref{Example_histogram}\,b).

\section{Finite-Size Scaling And Critical Behavior of The Localization Length\label{CriticalBehavior}}

\subsection{Results and Interpretation}

From Fig.\,\ref{Example_histogram}\,b) the localization length
is seen to increase with $\alpha$. Nevertheless, it is not evident
that it is diverging as predicted by Eq.\,\ref{LocalizationLenghtFormula}.,
due to the persistent finite-size scaling present in the data for
high values of $\alpha$.

\begin{figure}
\begin{raggedleft}
\includegraphics[width=8.62cm]{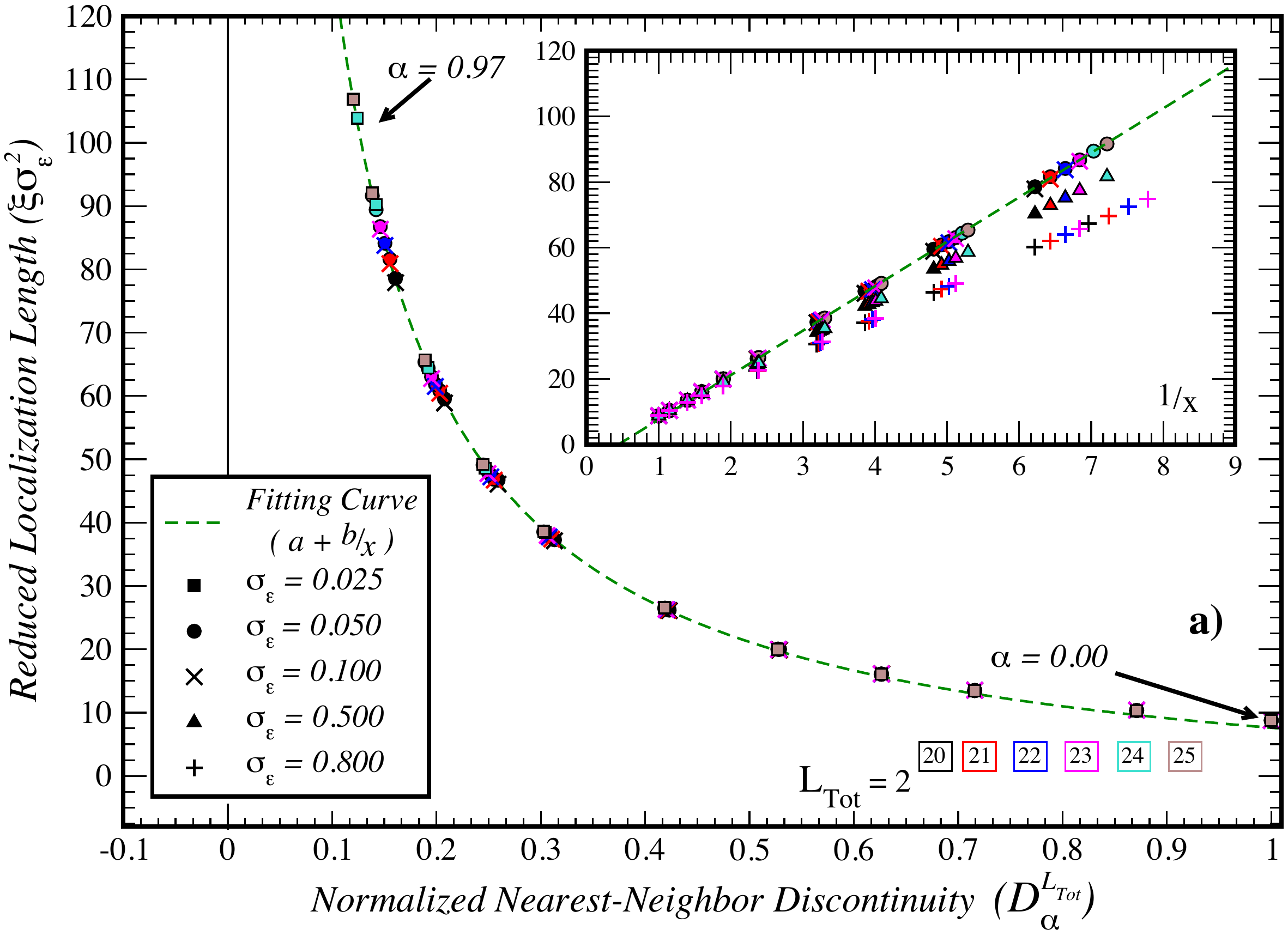}
\par\end{raggedleft}
\begin{raggedleft}
\includegraphics[width=8.5cm]{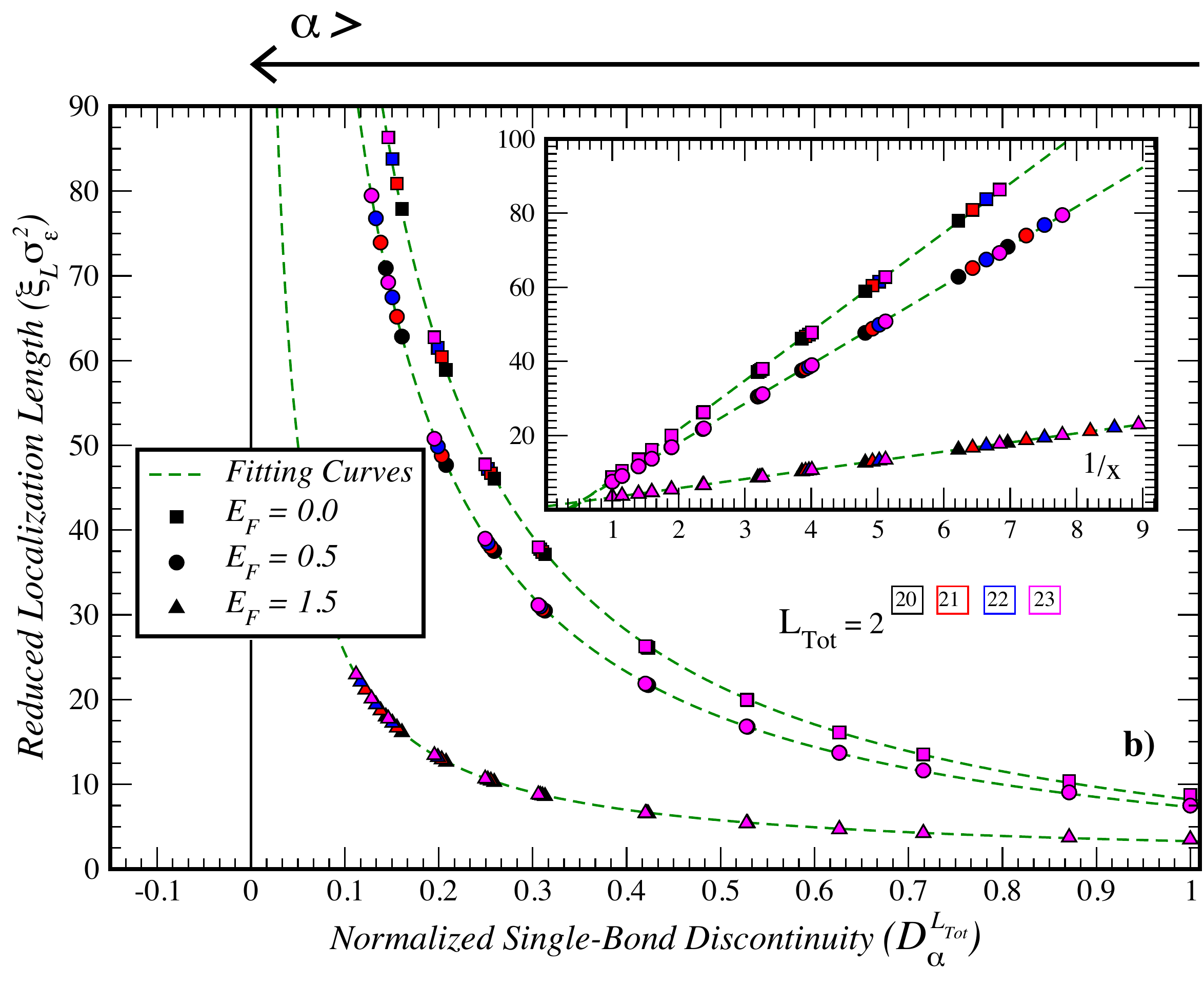}
\par\end{raggedleft}
\begin{raggedleft}
\includegraphics[width=8.61cm]{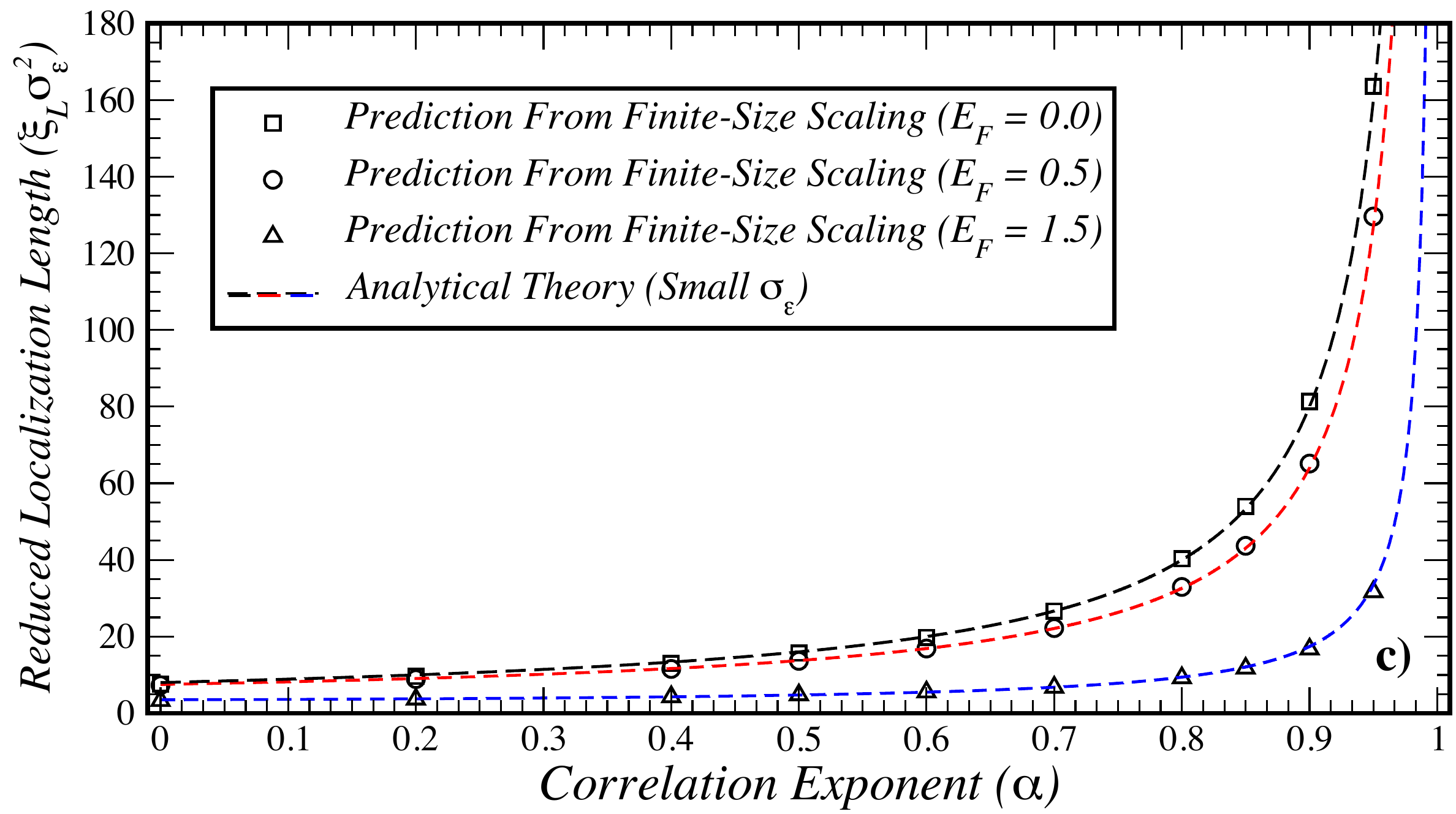}
\par\end{raggedleft}
\caption{\label{Collapsed_Data}\textbf{a) }Plot of\textbf{ }the localization
length as a function of the NSBD for different values $\sigma_{\varepsilon}$.
The green dashed curve corresponds to a fit for values close to $\mathcal{D}_{\alpha}^{L_{Tot}}=0$
of the sets with $\sigma_{\varepsilon}=0.025$, $0.05$ and $0.10$.
The data for $\sigma_{\varepsilon}=0.5$ and $0.8$ (only in the inset)
already show a deviation from this curve due to the breakdown of perturbation
theory. \textbf{b) }Plot of the localization length as a function
of the NSBD for different values of the energy and $\sigma_{\varepsilon}=0.10$.
All three cases can be collapsed into curves of the type $y=a+\frac{b}{x}$,
shown as dashed green lines.\textbf{ }The data is linearized by the
transformation $x\to\nicefrac{1}{x}$, in the insets.\textbf{ c)}
Comparison between the thermodynamic limit values of $\xi$ predicted
by the dashed curves in b), with the plots of the analytical expression
of Eq.\,\ref{LocalizationLenghtFormula}. (color online)}
 
\end{figure}

This scaling of the localization length is driven by the slow convergence
of the disorder statistics to its thermodynamic limit. In particular,
as referred in Sect.\,\ref{ModelHam_DisStat}B, even for values of
$\alpha$ very close to the transition point there is a sizable random
noise at small distances, whose amplitude goes very slowly to zero
as $L_{\text{Tot}}$ is increased. Our central argument in this paper
is that the main contribution to the eigenstates' localization comes
from this small-scale  component, which has an effective strength
measured by $\sqrt{\mathcal{D}_{\alpha}^{L_{\text{Tot}}}}$. The latter
claim can be proven by plotting the data in Fig.\,\ref{Example_histogram}\,b)
as a function of this parameter, instead of $\alpha$. This is done
in Fig.\,\ref{Collapsed_Data}\,a), and a perfect collapse of all
the points is obtained for small enough values of $\sigma_{\varepsilon}$.

The great advantage of this new representation is that we accomplish
a complete control over the finite-size scaling phenomenon: with increasing
in $L_{\text{Tot}}$, all the points slide over the dashed curve,
slowly approaching a fixed value. In other words, we turned a two-parameter
scaling, $\sigma_{\varepsilon}^{2}\xi=f_{1}\left(L_{Tot},\alpha\right)$,
into a single-parameter scaling, $\sigma_{\varepsilon}^{2}\xi=f_{2}\left(\mathcal{D}_{\alpha}^{L_{Tot}}\right)$.
This scaling law was used to extrapolate the values of $\xi$ to the
thermodynamic limit, with the following procedure:

\vspace{-0.2cm}
\begin{enumerate}
\item Calculate the thermodynamic limit values of $\mathcal{D}_{\alpha}^{\infty}=\lim_{L_{Tot}\to\infty}\left(\mathcal{D}_{\alpha}^{L_{Tot}}\right)$
from the Eqs.\,\ref{Correlator_ThermodynamicLimit} and \ref{NSBD},
for each value of $\alpha$;\vspace{-0.3cm}
\item Use the finite-size scaling curves (dashed curves in Fig.\,\ref{Collapsed_Data}\,b))
to read the values the thermodynamic limit of $\xi(\alpha)$;\vspace{-0.3cm}
\item Plot the corresponding values as a function of $\alpha$ and compare
them with the analytical expression of Eq.\,\ref{LocalizationLenghtFormula}.\vspace{-0.2cm}
\end{enumerate}
The final results are shown in Fig.\,\ref{Collapsed_Data}\,c)\,,
where it is clear that we have a perfect agreement with the analytical
results of Sect.\,\ref{Perturbative-Expression-for_LocLength}. For
completeness, we also checked this behavior for different energies,
which yielded a similar collapse of the data (see Fig.\,\ref{Collapsed_Data}\,b))
and agreement with Eq.\,\ref{LocalizationLenghtFormula}, thus confirming
the belief that this transition occurs over all the spectrum at once.
A direct fit of the data points in Fig.\,\ref{Collapsed_Data}\,c)\,,
to a function of the form $\xi\left(\alpha\right)=C\left(\alpha_{C}-\alpha\right)^{-\nu}$
yields the following results\footnote{The data set for $E_{F}=1.5$ do not have enough points close to the
singularity to provide a good estimate of $\nu$.}:
\begin{itemize}
\item For $E_{F}=0.0$: 
\begin{itemize}
\item $\alpha_{C}=1.0018\pm0.0064$;
\item $\nu=1.032\pm0.068$;
\end{itemize}
\item For $E_{F}=0.5$: 
\begin{itemize}
\item $\alpha_{C}=0.9970\pm0.0073$;
\item $\nu=0.953\pm0.075$;
\end{itemize}
\end{itemize}
which are in numerical agreement with the analytical expression.\vspace{-0.4cm}

\subsection{Delocalization or Insulator-Metal Transition?}

Before closing this section, it is imperative to make some further
comments on the physical interpretation of this divergence in $\xi$.
More precisely: Does this divergence signal a transition from an insulating
to a metallic phase? 

A first concern shall be raised about the applicability of our procedure
of cutting subchains from a larger sample and study their conductance,
when $\alpha\geq1$. Being non-stationary, the thermodynamic limit
correlator $\mathcal{C}_{\alpha}^{\infty}$ is a function of $\nicefrac{r}{L}_{Tot}$
(see Refs.\citep{Petersen2013,Khan2019}) and not simply of $r$,
as in Eq.\,\ref{Correlator_ThermodynamicLimit}. One consequence
of this is the following: If we take a fixed sized subchain from increasingly
larger chains, the values of $\varepsilon_{n}$ in the subchains will
become more and more correlated and, in the limit $L_{Tot}\to\infty$,
they will form an uniform potential inside the subchain. Hence, any
finite subchain will be ordered in this limit and therefore metallic.
Additionally, this argument invalidates immediately the existence
of a finite localization length for any $\alpha\geq1$.

However, one could have used subchains whose size is a fixed fraction
of $L_{Tot}$, i.e. $\gamma\equiv\nicefrac{L_{S}}{L_{Tot}}=\text{const}$.
In such a case, the reasoning of last paragraph is no longer valid,
but are the subchains still metallic? To answer this, we must remark
that in an insulator-to-metal transition, one has a diverging $\xi$,
but also a linear scaling of $\xi$ with the system size in the metallic
phase. As is known from previous work\citep{russ_localization_2001,BUNDE2000151},
the scaling of the localization length with the total system size
is anomalous\citep{BUNDE2000151} for any $\alpha>1$, i.e. $\xi\propto L_{Tot}^{\alpha-1}$.
An implication of this result is that the scaling of $\av{\log\frac{h}{e^{2}}G}$
is given by a function $f\left(\gamma,L_{Tot},\alpha\right)$, with
the following form:

\vspace{-0.3cm}

\begin{align}
f\left(\gamma,L_{Tot},\alpha\right) & =-\log\left[C_{1}\left(\alpha\right)\right]+\frac{L_{S}}{\xi\left(\alpha,L_{tot}\right)}\label{ModelAbovealfa1}\\
 & \qquad=-\log\left[C_{1}\left(\alpha\right)\right]+C_{2}\left(\alpha\right)\gamma L_{Tot}^{2-\alpha}.\nonumber 
\end{align}

\begin{figure}[H]
\includegraphics[scale=0.29]{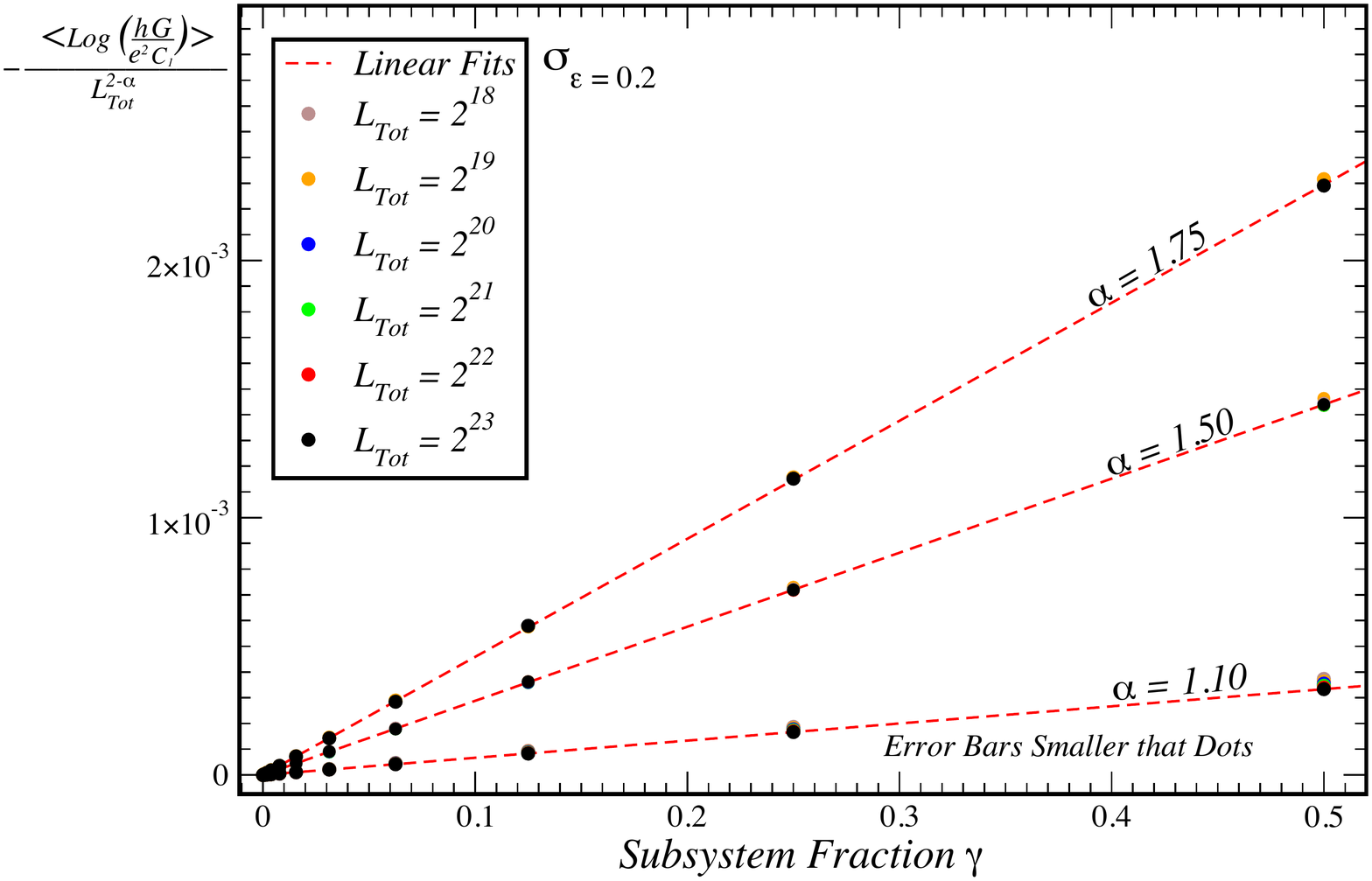}

\caption{\textbf{\label{Abov21}} Plots of the $\protect\av{\log\frac{hG}{e^{2}C_{1}\left(\alpha\right)}}$
as a function of the fraction chain used, $\gamma=L_{s}/L_{Tot}$,
for three values of $\alpha$ above the critical point. The $L_{Tot}$
dependence of the function $f(\gamma,L_{Tot,\alpha})$ in Eq.\,\ref{ModelAbovealfa1}
was used to collapse the data for different different system sizes
into a single straight line (dashed red lines). The value of the local
variance used was $\sigma_{\varepsilon}=0.2$.}
\end{figure}

For a start, Eq.\,\ref{ModelAbovealfa1} makes clear that for any
$\alpha>2$, the $\av{\log\frac{h}{e^{2}}G}$ converges to the finite
value $\log\left[C_{1}\left(\alpha\right)\right]$, in the limit $L_{Tot}\to\infty$
with $\gamma$ kept fixed. On the other hand, for $1\leq\alpha<2$
the situation is rather different, as it is the second term of Eq.\,\ref{ModelAbovealfa1}
that dominates in that same limit. Thus, the system becomes an insulator
as $L_{Tot}\to\infty$, as long as we take it with $\gamma$ fixed.
Both these facts explain the originally observed\citep{deMoura1998}
critical point at $\alpha=2$. At any rate, the scaling of Eq.\,\ref{ModelAbovealfa1}
may also be checked by calculations analogous to the ones we made
for $\alpha<1$. Our numerical results are shown in Fig.\,\ref{Abov21},
where we show the data of $\av{\log\frac{h}{e^{2}}G}\left(\gamma\right)$
for $6$ values of $L_{Tot}$ can be collapsed into a linear curve
after subtracting $C_{1}\left(\alpha\right)$ (obtained by fitting)
and rescaling by $L_{Tot}^{2-\alpha}$. This corroborates the validity
of Eq.\,\ref{ModelAbovealfa1} for $\alpha\geq1$.

Summarizing, the behavior of this model for $\alpha\to1^{+}$ is a
highly non-trivial one and, in particular, deciding whether or not
the phase is metallic is dependent on the way the thermodynamic limit
is taken. Therefore, what we mean by \emph{delocalization }in the
title of this paper is to be understood as the emergence of a phase
without a well-defined finite length scale $\xi$.

\section{Summary and Conclusions\label{Conclusions}}

In the previous sections, we discussed the problem of localization
in a correlated disordered potential, with a power-spectrum $\propto q^{-\alpha}$.
This model is known to have an ill-defined thermodynamic limit for
$\alpha\geq1$ (non-stationarity), but is perfectly well-defined for
$\alpha\in\left[0,1\right[$, having correlations decaying as $r^{\alpha-1}$
for long distances and a significant uncorrelation across a single
bond. This last feature generates a random  noise at short distances
with an effective strength given by the NSBD parameter \textemdash{}
$\sqrt{\mathcal{D}_{\alpha}^{L_{Tot}}}$. The finite-size effects
on the statistics of the disorder were also discussed and shown to
be very relevant near $\alpha=1$.

Furthermore, by applying a generalized Thouless Formula due to F.
M. Izrailev\,\citep{Izrailev1999}, we concluded that the localization
length in this system is expected to diverge as $\left(1-\alpha\right)^{-1}$
for $\alpha\rightarrow1^{-}$, at all points in the spectrum. These
results were confirmed by a direct measurement of the localization
length from the linear scaling of the typical Landauer conductance,
which was shown to suffer from a persistent finite-size scaling near
the transition point. This scaling was then related to the referred
finite-size effects in the statistics and the NSBD was found to be
the relevant parameter for studying it. By using the latter as a scaling
parameter for the localization length, we were able to collapse all
the data points into an universal curve, in the limit of small $\sigma_{\varepsilon}$.
With this picture, the finite-size scaling appear as a ``slide''
of the points along that curve, approaching their thermodynamic limit
value for each $\alpha$. Finally, the limiting curve was shown to
be consistent with the result from Izrailev's formula, and the existence
of a global delocalization transition at $\alpha=1$.

To finish, we summarize our main conclusions in two two points: 
\begin{enumerate}
\item We were able to numerically observe the existence of a delocalization
transition in these models, at $\alpha=1$, and show that it agrees
with analytical results in the weak-disordered regime. To the best
of our knowledge, this is a novelty in the literature, which settles
the matter in favor of an expected global delocalization of the eigenstates
for $\alpha\geq1$;
\item The way we were able to control the finite-size scaling of the localization
length provides us with very clear evidences about the true nature
of the said transition. As a matter of fact, the observed behavior
of $\xi\left(\mathcal{D}_{\alpha}^{L_{Tot}}\right)$ in the perturbati\textcolor{black}{ve
regime is exactly the same as one would expect from an uncorrelated
Anderson disorder, with an effective strength given by $\sqrt{\mathcal{D}_{\alpha}^{L_{Tot}}}.$
This fact gives a strong indication that the variation of the localization
length with the exponent $\alpha$ is mainly due to a varying effective
strength of the short-scale noise, and not due to the change in the
tail's exponent of the real-space disorder correlator. Nonetheless,
the two effects cannot be decoupled in the de Moura-Lyra model as
there is there is a single parameter, $\alpha$, controlling both
the vanishing of the local noise component and the power-law tail's
exponent.}\vspace{-0.2cm}
\end{enumerate}
In our perspective, both points are equally relevant for the physical
interpretation of this disorder model, in the sense that both lead
to a final conclusion: the model is as unable to generate 1D Anderson
transitions (for which it was built), as it is of reproducing the
effects power-law tails in the real-space disorder correlator. \textcolor{black}{In
fact, the localization phenomena seems to be as simple here, as in
an uncorrelated Anderson model and the delocalization transition is
a disorder-to-order one. An interesting follow-up to the present work
would be finding new ways of simulating stationary disorder landscapes
having power-law tails with tunable exponents }but a possibly negligible
and independently determined small-scale noise\textcolor{black}{.
Analyzing systems with such potentials would give us information about
the importance of algebraic correlation tails in the physics of one-dimensional
localization.}

\section{Acknowledgments}

We thank the helpful comments and suggestions of the Physical Review
B referees, which contributed to the improvement of this work.

For this work, J. P. Santos Pires was supported by the MAP-fis PhD
grant PD/BD/142774/2018 of Fundação da Ciência e Tecnologia; N. A.
Khan was supported by the grants ERASMUS MUNDUS Action 2 Strand 1
Lot 11, EACEA/42/11 Grant Agreement 2013-2538 / 001-001 EM Action
2 Partnership Asia-Europe and the research scholarship UID/FIS/04650/2013
of Fundação da Ciência e Tecnologia. The authors also acknowledge
financing from Fundação da Ciência e Tecnologia and COMPETE 2020 program
in FEDER component (European Union), through the projects POCI-01-0145-FEDER-02888
and UID/FIS/04650/2013.\vspace{4.0cm}

\bibliographystyle{apsrev4-1}
\bibliography{References}

\end{document}